# HumanErr, SSR: 2003.

# Reducing Overconfidence in Spreadsheet Development

Dr. Raymond R. Panko
University of Hawaii
College of Business Administration
2404 Maile Way
Honolulu, HI 96822

Ray@Panko.com
Panko@Hawaii.edu

**ABSTRACT**

Despite strong evidence of widespread errors, spreadsheet developers rarely subject their spreadsheets to post-development testing to reduce errors. This may be because spreadsheet developers are overconfident in the accuracy of their spreadsheets. This conjecture is plausible because overconfidence is present in a wide variety of human cognitive domains, even among experts. This paper describes two experiments in overconfidence in spreadsheet development. The first is a pilot study to determine the existence of overconfidence. The second tests a manipulation to reduce overconfidence and errors. The manipulation is modestly successful, indicating that overconfidence reduction is a promising avenue to pursue.

**1. INTRODUCTION**

Spreadsheeting was one of the earliest end user applications, along with word processing. It continues to be among the most widely used computer applications in organizations [United States Bureau of the Census, 1999]. Although many spreadsheets are small and simple throwaway calculations, surveys have shown that many spreadsheets are quite large [Cale 1994; Cragg & King 1993; Floyd, Walls, & Marr, 1995; Hall 1996], complex [Hall, 1996], and important to the firm [Chan & Storey, 1996; Gable, Yap, and Eng, 1991].

**1.1 Evidence of Errors in Spreadsheet Development**

Unfortunately, there is growing evidence that incorrect spreadsheets are commonplace. For instance, Table 1 shows that audits of real-world spreadsheets have found errors in many of the spreadsheets they examined. The four most recent studies, which used better methodologies than earlier studies, found errors in 91 % of the 54 spreadsheets they inspected. In the laboratory, in turn (see Table 1), uncorrected errors were found in at least one percent of the formula cells of the spreadsheets built during experiments [Panko, 2001b]. If there are errors in anything like this percentage of cells in real world spreadsheets, the implications are sobering. As Table 1 shows, the few field audits that measured the frequency of errors on a per-cell basis indicate that undetected error rates really are about this size [Butler, 2000; Hicks, 1995; Lukasic, 1998].

*Table 1: Studies of Spreadsheet Errors (see Tables and Figures)*

Although such uncorrected error rates are troubling, they should not be surprising. Human error research has shown consistently that for nontrivial cognitive actions, undetected and therefore uncorrected error rates of a few percent are nearly always present [Panko, 2001 a]. In software development, for instance, over 20 field studies have shown that a few percent of all lines of code will always be incorrect even after a module is carefully developed.

### 1.2 Infrequent Post-Development Testing

In the face of such high error rates, software development projects usually devote about a third of their effort to post-development error correction [Grady, 1994; Jones, 1998]. Even after several rounds of post-development testing, some errors remain [Putnam & Myers, 1992].

The picture in end user computing, however, is very different. Organizations rarely mandate that spreadsheets and other end user applications be tested [Cale, 1994; Cragg & King, 1993; Floyd, Walls, & Marr, 1995; Galletta & Huffnagel, 1992; Hall, 1996; Speier & Brown, 1996], and individual developers rarely engage in systematic testing on their own [Cragg & King, 1993; Davies & lkin, 1987; Hall, 1996; Schultheis & Sumner, 1994].

### 1.3 Overconfidence in Spreadsheet Development

Why is testing so rare in spreadsheet development in the face of substantial error rates in spreadsheeting and substantial error rates in other human cognitive domains? The answer may be that spreadsheet developers are overconfident of the accuracy of their spreadsheets. For instance, Brown and Gould [1987] had nine highly experienced spreadsheet developers each create three spreadsheets, from word problems. Sixty-three percent of the 27 spreadsheets developed contained errors, and all of the nine developers made at least one error. Yet, when subjects were asked to rate their confidence in the accuracy of their spreadsheets, their mean response was "quite confident." High confidence in the correctness of spreadsheets has also been seen in other spreadsheet experiments [Panko & Halverson, 1997], field audits [Davies & Ikin, 19871, and surveys [Floyd, Walls, & Marr, 19951.

However, these measurements of spreadsheet overconfidence have used 5-point and 7-point Likert scales. These can be difficult to interpret. For instance, when developers in the Brown and Gould's [1987] experiment rated themselves as "quite confident," this was still only four on a scale of five, which perhaps indicates only moderate confidence.

### 1.4 This Paper

This paper presents two experiments to shed light on overconfidence in spreadsheet development. The first uses a more easily interpreted measure of overconfidence to address whether high levels of confidence seen in Likert scales really are as extreme as they seem to be. Specifically, each subject was asked to estimate the probability that he or she had made an error during the development of their spreadsheet. The mean of these estimates was compared with the actual percentage of incorrect spreadsheets.

The second study used a manipulation to see if the manipulation could reduce overconfidence and, hopefully, improve accuracy as a consequence of reduced overconfidence. Specifically, subjects in the treatment group were told the percentage of subjects who had produced incorrect spreadsheets from the task's word problem in the past, while subjects in the control group were not given this information. Both groups then built spreadsheets from the word problem.

### 2. OVERCONFIDENCE

In studying overconfidence, we can draw upon a broad literature. Overconfidence appears to be a general human tendency. Research has shown that most people believe that they are superior to most other people in many areas of life [Brown, 1990, Koriat, Lichtenstein, & Fischoff, 1980]. Nor is overconfidence limited to personal life. Problem solvers and planners in industry also tend overestimate their knowledge [Koriat, Lichtenstein, & Fischoff, 1980]. Indeed, a survey of the overconfidence literature [Pulford & Colman, 1996] has show that overconfidence is one of the most consistent findings in behavioral research.

## 2.1 The Danger of Overconfidence

Overconfidence can be dangerous. In driving, one of the main causes of accidents is the late detection of errors because overconfident drivers do not take avoidance measures until it is too late to recover [Rumar, 1990]. In error detection, we have stopping rules that determine how far we will go to look for errors [Rasmussen, 1990]. If we are overconfident in our accuracy, we may stop looking for errors too soon. If spreadsheet developers are overconfident, this may lead them to stop error detection short of formal testing after development.

## 2.2 Risky Behaviour can Reinforce Overconfidence

Fuller [1990] noted that engaging in risky behaviour actually can be self-reinforcing. If we take risky actions when we drive, this rarely causes accidents, so we get little negative feedback to extinguish our behaviour. At the same time, if we speed, we arrive earlier, and this reinforces our risky behavior. In spreadsheeting, developers who do not do comprehensive error checking are rewarded both by finishing faster and by avoiding onerous testing work.

Even if we find some errors as we work, this may only reinforce risky behaviour. In a simulation study of ship handling, Habberley, Shaddick, and Taylor [1986], observed that skilled watch officers consistently came hazardously close to other vessels. In addition, when risky behaviour required error avoiding actions, the watch officers experienced a gain in confidence in their "skills" because they had successfully avoided accidents. Similarly, in spreadsheeting, if we catch some errors as we work, we may believe that we are skilled in catching errors and so have no need for formal testing.

## 23 The Hard-Easy Effect

The most consistent finding within laboratory overconfidence research is the "hard-easy effect" [Clarke, 1960; Lichtenstein, Fischoff, & Philips, 1982; Plous, 1993; Pulford & Coleman, 1996; Wagenaar & Keren, 1986]. In studies that have probed this effect, subjects were given tasks of varying difficulty. These studies found that as accuracy fell in more difficult tasks, confidence levels fell only slightly, so that overconfidence increased. Given the high number of errors found in spreadsheet audits and experiments, spreadsheeting appears to be a difficult task. Accordingly, we would expect to see substantial amounts of overconfidence in spreadsheet development.

## 2.4 Manipulating confidence

Several procedural innovations have been tried to reduce overconfidence. One study [Lichtenstein & Fischoff, 1980] found that systematic feedback was useful. Over a long series of trials, subjects were told whether they were correct or not for each question. Overconfidence decreased over the series. In another study [Arkes, et al., 1987], subjects lost their overconfidence when given feedback after five deceptively difficult problems. In addition, we know from Kasper's [1996] recent overview of DSS research that merely providing information is not enough. Feedback on the correctness of decisions must be detailed and consistent. These studies collectively suggest that feedback about errors can reduce overconfidence.

## 2.5 Overconfidence Among Experts

Most laboratory studies, like the ones described in this paper, use students as subjects. However, studies have shown that experts also tend to be overconfident when they work [Shanteau & Phelps, 1977; Wagenaar & Keren, 1986]. In fact, when experts find themselves wrong, they often believe that the sources of inconsistencies lie elsewhere than in themselves [Shanteau & Phelps, 1977].

One puzzle from research on experts is that experts in some occupations are very well calibrated in confidence [Keren, 1992; Shanteau & Phelps, 1977; Wagenaar & Keren, 1986], while in other occupations they are very poorly calibrated [Camerer & Johnson, 1991; Johnson, 1988; Shanteau & Phelps, 1977; Wagenaar & Keren, 1986]. Shanteau [1992] analyzed situations in which experts were either well or poorly calibrated. He discovered that experts tend to be well calibrated only if they receive consistent and detailed feedback on their error rates. Wagenaar and Reason [1990] also emphasized the importance of experts comparing large numbers of predictions with actual outcomes in a systematic way if their confidence is to be calibrated. This need for analyzed feedback among professionals is reminiscent of results from laboratory research to reduce overconfidence noted earlier.

Note that experience is not enough. Many studies of experts looked at people with extensive experience. In many cases, however, these experts did not receive detailed and consistent feedback. For instance, blackjack dealers, who merely deal and have no need to analyze and reflect upon the outcome of each deal afterward, are not better calibrated than lay people at blackjack [Wagenaar & Keren, 1986]. In contrast, expert bridge players get feedback with each hand and analyze that feedback in detail [Wagenaar & Keren, 1986]. They are well calibrated in confidence.

As noted above, spreadsheet developers rarely audit their spreadsheets in detail after development. With little systematic feedback because of the rarity of post-development testing, it would be surprising if spreadsheet developers were well-calibrated in their confidence. In contrast, one of the tenets of software code inspection is the reporting of results after each inspection [Fagan, 1976]. Therefore, software developers, who do extensive testing and also get detailed feedback for analysis, have the motivation to continue doing extensive testing because of the errors this testing reveals.

**2.6 Overconridence in Groups**

Most overconfidence studies have looked at individuals. However, managers spend much of their time in face-to-face meetings. Therefore, we would like to know if groups, like individuals, are chronically overconfident. Despite the focus of most overconfidence research on individuals, there is evidence that overconfidence also occurs in group settings [Ono & Davis, 1988; Sniezek & Henry, 1989; Plous, 1995].

**2.7 Conclusions Regarding Spreadsheeting**

Although the overconfidence literature is largely empirical and is weak in theory, a number of research results suggest that overconfidence is an important issue for spreadsheet accuracy.

- First, the broad body of the literature has shown that overconfidence is almost universal, so we should expect to see it in spreadsheeting.

- Second, overconfidence tends to result in risky behavior, such as not testing for errors.

- Third, error rates shown in Table 1 indicate that spreadsheeting is a difficult task, so in accordance with the hard-easy effect, we should expect substantial overconfidence.

- Fourth, even experts are poorly calibrated in confidence unless they do consistent and reflective analysis after each task, which is uncommon in spreadsheeting.

- Fifth, it may be possible to reduce overconfidence by providing feedback.

- Sixth, reducing overconfidence may reduce errors, although this link is not demonstrated explicitly in the overconfidence literature.

## 3. EXPERIMENT 1: ESTABLISHING THE PRESENCE OF OVERCONFIDENCE

In our first experiment, the goals were simple: to see if high confidence ratings seen previously with Likert scale questions really indicated a very low perceived likelihood of making an error, and to see if the method for measuring confidence used in this study appears to be useful. We measured confidence after development, and we had no manipulation of confidence.

### 3.1 Sample

The sample consisted of upper-division undergraduate management information systems majors in the business school of a medium-size state university. All had taken a course that taught spreadsheeting and a subsequent course that used spreadsheets extensively. Subjects engaged in the experiment to receive extra credit. Over 80% of the students in the class participated. Accounting and finance students were excluded because of their specialized task knowledge. This left 80 participants. Subjects were randomly assigned to work alone or in groups of three (triads). Forty-five students were assigned to triads, while thirty-five developed the spreadsheet working alone.

### 3.2 Task

The task used in this study was the MicroSlo task, which required students to build a pro-forma income statement from a word problem:

Your task is to build a two-year pro forma income statement for a company. The company sells microwave slow cookers, for use in restaurants. The owner will draw a salary of $80,000 per year. There is also a manager of operations, who will draw a salary of $60,000 per year. The income tax rate is expected to be 25% in each of the two years. Each MicroSlo cooker will require $40 in materials costs and $25 in labor costs in the first year. These numbers are expected to change to $35 and $29 in the second year. Unit sales price is expected to be $200 in the first year and to grow by 10% in the second year. There will be three sales people. Their salary is expected to average $30,000 in the first year and $3 1,000 in the second. Factory rent will be $3,000 per month. The company expects to sell 3,000 MicroSlo cookers in the first year. In the second, it expects to sell 3,200.

MicroSlo is a modification of the Panko and Halverson [1997] Galumpke task. Wording was changed to avoid potential confusion, and the capital purchase subtask, which most students could not do successfully, was removed.

### 3.3 Dependent Variables

After subjects had built the spreadsheet, they wore asked to estimate the probability that they (or their triad) had made an error when building the spreadsheet. Subjects also estimated the percentage of individuals and triads other than themselves that had made an error. Likelihood estimates could vary from 0% to 100%. Likelihood of error estimates were treated as interval variables in the data analysis.

### 3.4 Procedure

Subjects were randomly assigned to the two conditions (working alone or in triads). Subjects working alone used computers in a common room. They were monitored to prevent cheating. Triads worked in other rooms, one triad to a room. Each triad had a single computer.

## 3.5 Results

As Table 2 shows, subjects were highly overconfident, based on our scale for estimating errors. For subjects working alone, the mean expected probability of an error in their own spreadsheet was only 18%. In fact, 86% of the subjects working alone made errors. For triads, expected errors in their own spreadsheet were "only" about half-13% instead of 27%. As research on the hard-easy effect would lead us to expect, the better calibration of triads was more due to better performance than to reductions in confidence. Expected error frequencies for other individuals and triads working at the task were also poorly calibrated.

*Table 2: Overconfidence in Monadic and Triadic Spreadsheet Development (see Tables and Figures)*

To test whether miscalibration was statistically significant, we used the standard but imperfect test [Dunning, Griffin, Milojkovic, & Ross, 1990] of comparing the mean of the predicted probability of error distribution with the actual frequency of error for the sample (86% for individuals and 27% for triads). Specifically, we tested whether the actual frequency was within the 95% confidence interval of the predicted mean.

Table 2 shows that only the estimation of errors for other triads failed to reach the 0.05 cutoff for significance, and it came very close. For all other categories, the difference was highly significant.

Another way to see miscalibration is to look at the percentage of estimates that were below the group mean. Table 2 shows that in three of the six estimates, all subjects had an expected probability of error lower than the group mean. The lowest percentage of individual scores under the group mean was 80%. Overconfidence, in other words, seemed to be pervasive.

Finally, in line with other research on overconfidence [Brown, 1990], subjects working alone felt that other subjects working alone would do worse than they would. Subjects working in triads likewise thought that their triad would do better than other triads.

## 3.6 Discussion

This experiment showed that subjects were indeed overconfident, substantially underestimating the probability that they had made an error and also believing that their odds of making an error were less than those of others. Our subjects, in other words, exhibited classic overconfidence behavior.

In fact, this overconfidence was persistent. After the experiment, when one class was debriefed, they were shown the data and were amazed by the results. When subjects who had worked alone were asked to raise their hand if they thought they were one of the 18% that did the task correctly, well over half raised their hands.

## 4. EXPERIMENT II: WILL WARNINGS HELP?

As noted earlier, feedback may be able to reduce overconfidence [Arkes et al., 1987; Kasper, 1996; and Liechtenstein & Fischoff, 1980]. Ideally, a person should receive feedback on their own development errors over a series of trials. However, perhaps simply giving the person an indication of how many people had made errors previously in specific circumstances may be able to help. To explore this conjecture, the second experiment had subjects develop two spreadsheets from word problems and attempted to manipulate their overconfidence by telling them the percentage of previous subjects that had made errors in the task.

## 4.1 Research Model

Figure 1 shows our research model for confidence and accuracy in spreadsheet development. This model is based on stages of development. For each stage, we are concerned with degree of accuracy, degree of confidence, and the interaction of accuracy and confidence.

*Figure 1: Research Model (see Tables and Figures)*

**Initial Confidence**

In Experiment 2, we manipulate initial (pre-development) confidence by giving the subject information about the percentage of past subjects who have made errors on the two tasks used in the experiment. This is a surrogate for the feedback that developers would have if they did postdevelopment testing and analyzed the results systematically as is done in software code inspection to provide more realistic knowledge about error rates [Fagan 1976]. This leads to our first hypothesis, HI:

> H1: Initial confidence should depend on the manipulation of telling the treatment subjects the percentage of past subjects who had created incorrect spreadsheets from this word problem.

**During Development**

Allwood. [1984] and others have observed human problem solving systematically. They have shown that people detect and correct many of the errors they make during development. Olson and Nilsen [1987-1988] specifically noted error detection and correction during spreadsheet development. However, as noted above, when people detect and successfully correct errors, this may increase their confidence. This leads to hypothesis H2:

> H2: Confidence after development should be higher than initial confidence before development.

**After Development**

We assume that the effect of the manipulation will last through the development stage. Consequently, we expect confidence to be lower in subjects who received the manipulation than in the control group.

> H3: The manipulation should make confidence lower in the treatment group than in the control group after development.

As noted in the first experiment, subjects have expectations about how many errors other subjects would make working on the task. We expect projected confidence to be lower in subjects who received the manipulation than in the control group.

> H4: The manipulation should make expected post-development confidence for other subjects lower in the treatment group than in the control group after development.

**Accuracy**

Finally, although the literature does not give us a direct link between confidence and accuracy, we assume that reducing confidence by manipulation (or, in the real world, by systematically analyzing post-development testing results) should increase accuracy. Otherwise, why bother? This leads to our final hypothesis:

> H5: Subjects who receive the manipulation of being told the percentage of past students who have created incorrect spreadsheets from this task should have a larger percentage of accurate spreadsheets.

**During and After Testing**

Testing is rare in spreadsheet development, as noted earlier. However, if testing were done, the developer would be almost certain to find undetected errors in his or her "clean" spreadsheet. This would increase the accuracy of the spreadsheet. In addition, his or her confidence should fall as objective proof of errors is seen. If testing and analysis of the results is done systematically, confidence should be better calibrated, as indicated by research on feedback and calibration discussed earlier. This study does not address the testing phase.

**4.2 Sample**

The sample consisted of upper-division (third and fourth year) management information systems majors in the college of business administration in a medium-size state university. All had taken a course that taught spreadsheeting and a subsequent course that used spreadsheets extensively. All subjects were taking a management information systems course. They received extra credit for participating in the study. Participation was voluntary. Over 80% of the class participated. Sixty students participated in the experiment, but five data sets had to be discarded because the student. failed to fill out the confidence survey information (three data sets) or because the files on disk could not be read (two data sets). This resulted in 55 useful data sets-27 in the control group and 28 in the treatment group.

**4.3 Tasks**

All students completed two tasks, which they performed in random order. One was the Kooker task, which was the MicroSlo task shown in the previous experiment with a capital purchase added. The other was the Wall task, developed by Panko and Sprague [1998]

You are to build a spreadsheet model to help you create a bid to build a wall. You will offer two options-lava rock or brick. Both walls will be built by crews of two. Crews will work three eight-hour days to build either type of wall. The wall will be 20 feet long, 6 feet tall, and 2 feet thick. Wages will be $ 10 per hour per person. You will have to add 20% to wages to cover fringe benefits. Lava rock will cost $3 per cubic foot. Brick will cost $2 per cubic foot. Your bid must add a profit margin of 30% to your expected cost.

**4.4 Dependent Variables**

Subjects were asked to estimate the probability that they would make an error building each of their spreadsheets after reading the task statement and after a warning (for the treatment group) but before doing the task. Estimates could range from 0% to 100%. We call this probability the *expected error probability (EEP).* In the analysis, we treated it as in interval variable.

Higher values for estimated likelihoods of making an error indicate lower confidence. We will use this relationship throughout the analysis.

For the accuracy measure, if bottom-line values were equal to those in a standard solution, the spreadsheet was deemed to be correct. Otherwise, it was deemed incorrect. We decided not to use the number of errors as our error estimate because the error distribution was highly skewed with a strong zero and a long tail. We could not find a reasonable way to handle normalization problems or the treatment of many zero values. In addition, some spreadsheets were wildly incorrect. For instance, three subjects produced Kooker solutions that looked nothing like income statements. There was no way to count their number of errors, and excluding these spreadsheets presented conceptual difficulties as well.

**4.5 Procedure**

After reading the description of the task, subjects in the treatment group were told in writing the percentage of subjects who had made errors doing this task in the past (80% for Kooker and 40% for Wall). This information was written in boldface, and the experimenter emphasized verbally that some subjects had boldface information and should read it carefully while other subjects did not have such information. After the experiment, a dozen subjects were asked if they had seen the boldface information and to characterize it. The seven who should have seen the information in their packet all characterized it correctly as indicating how many people had made errors in the past. The five who could not have seen the information all said that they had not seen boldface information.

All subjects worked in a computer laboratory. They were monitored as they worked to ensure that there was no discussion and that students were not watching other screens.

### 4.6 Results

Table 3 shows the results for estimated error probabilities (EEPs) and spreadsheet correctness. In the results, note that a higher EEP indicates less confidence than a lower EEP. Also, except in H2, the hypothesis testing was done on the average EEP of the subject across both tasks; the two EEPs were added and divided by two.

*Table 3: Results of the Warning Experiment (see Tables and Figures)*

**Overconfidence Before Development**

For Hypothesis HI, the expected error probability was measured before development. For the Kooker task, the difference in average EEP was substantial: 40% in the control group versus 60% in the treatment group. For the Wall task, improvements were again encouraging. The two values were 34% to 53%, respectively. For the EEP averaged across the two tasks, the values for the two groups were 37% for the control group and 56% for the treatment group. As expected, EEP was higher in the treatment group than in the control group, indicating that the manipulation did reduce confidence.

This difference in EEP averaged across the two tasks was significant at the 0.008 level (t=2.48, df=52). Consequently, we reject the null hypothesis and accept H1: that giving a warning reduces confidence before development. This is an important finding, because if it were not true, the entire experiment would be a failure.

**Decline in Confidence After Development**

As noted earlier, we expected that subjects would detect and correct some errors during development, and this should increase their overconfidence. According to H2, confidence should increase after development. We measured this difference with a paired t-test. All subjects were used in the test because the sample size was too small to study interaction effects.

The average EEP for the two tasks was 49% before development. This fell to 3 8% after development, indicating the expected increase in confidence. The difference is statistically significant beyond the 0.0000 level (t=4.58, df=50). Consequently, we reject the null hypothesis and accept H2: that confidence increases after development.

**Overconfidence After Development (Subject)**

H3 expects the manipulation to remain effective after development. As Table 3 shows, the EEP averaged over both tasks after development was 3 1% on average in the control group and 43% on average in the treatment group. Although this difference was in the expected direction, it was not statistically significant, only reaching a probability of 0.070 (t= 1.50, df=51). Hypothesis 3 was not supported.

**Overconfidence After Development (Other)**

H4 focuses on how subjects believe others would do on the tasks they completed. As Table 3 shows, the subjects estimated the average EEP of others across both tasks as 23% on average in the control group and 40% on average in the treatment group. This difference was significant at the 0.001 level (t=3.27, df=52). Consequently, we reject the null hypothesis and accept H4, that confidence is lower in the treatment group than in the control group.

**Correctness**

The percent of correct spreadsheets was taken as a proportion variable. A z-test for proportions was used to determine whether the difference in percentages between the experimental and control treatments was statistically significant. The dependent variable was whether both spreadsheets were correct or not.

In the control group, only 7% of all spreadsheets were correct. However, in the treatment group, 25% of the spreadsheets were correct. Of course, the improvement looks smaller if you compare the percentage of incorrect spreadsheets-93% and 75%, respectively. The difference, while not large enough to make spreadsheet development safe, was statistically significant at the 0.40 level (z--1.78).

### 4.7 Discussion

As expected, our subjects were again broadly overconfident. While the percentage of incorrect spreadsheets in the control group and treatment group were 93% and 73%, respectively, not one of the estimated error probabilities was anywhere near this high.

Second, as expected, confidence rose during the development phase. This provides support for the theory that finding errors during development increases confidence.

Third, the manipulation-information about the rates of incorrect spreadsheets for previous subjects-tended to reduce overconfidence. The decrease was statistically significant for the initial EEP and for the expected EEP for other developers. It was not statistically significant for the developers' own EEP after development, although it was in the expected direction.

Fourth, and perhaps most importantly, the manipulation increased the percentage of correct spreadsheets from 7% in the control group to 25% in the treatment group. This is encouraging because it represents more than a tripling in the percentage of spreadsheets that were correct. However, it does not make spreadsheet development safe.

Unfortunately, the experiment could not measure whether subjects were more likely in the future to engage in systematic testing after development. Unless such full testing is done, error rates must be expected to remain unacceptably high.

## 5. CONCLUSION

### 5.1 Confidence and Accuracy

As noted in the Introduction, overconfidence is an almost universal human trait. The first experiment demonstrated that spreadsheet development is no exception. On average, subjects working alone rated the probability that they had made an error as only 18%. In fact, 86% made errors. Subjects working alone also believed that other subjects working alone were more likely to have made an error than they themselves were. Results for subjects working in triads were less extreme but still exhibited classic overconfidence patterns.

Can we reduce overconfidence, and if we do, will accuracy increase? The second experiment gives a cautious "yes" in response to these questions. Not all decreases in confidence were statistically significant, but all were in the expected direction. More importantly, reducing confidence increased accuracy. With a warning, the percentage of subjects getting both spreadsheets correct more than tripled, from 7% to 25%.

One point of caution in interpreting the results is that even with the improvements seen when warnings were given, subject accuracy still was too low to make spreadsheet development a safe activity. Unless feedback can decrease overconfidence sufficiently to motivate users to test their spreadsheets systematically, we are unlikely to get the order-of-magnitude error reduction needed to make spreadsheets even marginally safe.

**5.2 The Need for a True Feedback Study**

Given the results of this experiment, a logical next step would be to follow the approaches of Arkes, *et al.* [1987] and Lichtenstein and Fischoff [1980] and give subjects a series of spreadsheet development tasks, providing them with detailed feedback on errors at each step.

**5.3 Wording and Statistical Thinking**

In addition, it would be useful in future studies to see if wording could affect confidence estimates. In this experiment, we asked subjects to estimate the likelihood that they would make an error. It would be interesting to see if compatible data would result if we asked subjects to estimate the likelihood that their spreadsheet was correct.

More fundamentally, we need to understand how subjects think about accuracy and the multiplication of probabilities. Subjects were asked in the second experiment to estimate the likelihood that they would make an error in a single cell, rather than for the spreadsheet as a whole. Their average percent estimate for a cell was 12.5%, and there was almost no difference between the two groups. This estimate of 12.5% is shockingly low because there were 42 cells in the model solution for Kooker and 25 in the model solution for Wall. If there are N formula cells in a spreadsheet, and if the probability of making an error in a cell is e, the probability of an error in at least one of the cells should be $E = 1 - (1+e)^N$ [Lorge & Solomon, 1955]. With a cell error rate of 12.5%, the probability of Kooker being incorrect is 99.6%. The probability of Wall being incorrect would be 96.5%. These high error rates are inconsistent with both subject estimates and actual spreadsheet error rates. In other words, subjects have inconsistent estimates for cell error rates and for the probability that they would make an error in their spreadsheet. This indicates that lay statistical thinking may lie at the root of overconfidence.

This misunderstanding of how a long series of calculations increases the likelihood of a error also seems to be indicated in a study by Reithel, *et al.* [1996], who had subjects look at spreadsheets that were short and poorly-formatted, short and well-formatted, long and poorly-formatted, and long and well-formatted. Subjects expressed substantially more confidence in the accuracy of the long and well-formatted spreadsheet than in the accuracy of other spreadsheets, despite the fact that longer spreadsheets should have more errors than shorter spreadsheets, given the multiplication of probabilities.

One surprise in the second experiment was that subjects on average thought that other developers would have *lower* error rates than they themselves would. This is the opposite of what one would expect from the general finding, mentioned earlier, that people tend to think more highly of themselves than others. It is also the opposite of what we saw in the first experiment. This also should be probed in future studies.

**5.4 The Use of Undergraduate Students**

One limitation in the experiment was the use of undergraduate subjects. Although all had previously taken a hands-on skills course and had undertaken at least two spreadsheet development homework assignments in their current class, none had extensive spreadsheet development experience at work. However, Panko and Sprague [1998] found almost identical error rates for the Wall task when the task was solved by undergraduate students, NWA students with little or no spreadsheet development experience at work, and MBA students with extensive spreadsheet development experience at work. Also, when Galletta *et al.* [1993] conducted a study in spreadsheet auditing (error detection through code inspection), the only difference between 11BA students with little or no development experience at work and those with more than 250 hours of development experience at work was speed of task completion. There was no significant difference in error detection rates.

**5.5 Theory**

For the future, a major task for the overconfidence literature must be to go beyond its impressive body of empirical results and move to theory creation. Although we cannot offer a full theory, we offer several suggestions for directions such a theory may take.

Reason [1990] and Baars [1992] have argued that human cognition has two mechanisms that differ in important characteristics. First, we have an automatic cognition system that uses pattern matching. This automatic system is fast and effortless [Reason. 1990]. Second, we have an attentional system that is linear, slow, and effortful. Testing, for instance code inspection, appears to require a high degree of effort and tends to be unpleasant for people [Beck, 20001. Human beings appear to be have a difficult time engaging the attentional system for more than brief periods of time. This may create strong resistance to formal testing.

There is a literature on denial, which focuses on illness and the fact that many people with terminal illnesses deny the seriousness of their condition or the need to take action. Apparently, what is very difficult and unpleasant to do is difficult to contemplate. Although denial has only been studied extensively in the medical literature, it is likely to appear whenever required actions are difficult or onerous. Given the effortful nature of spreadsheet testing, developers may be victims of denial, which may manifest itself in the form of overconfidence in accuracy so that extensive testing will not be needed.

Reinforcing the denial possibility for explaining overconfidence is the study by Reithel *et al.* (1996) mentioned earlier. For large well-formatted spreadsheets, expressed confidence was much higher than for shorter spreadsheets. One explanation may be that subjects seeing a spreadsheet too large to test easily, and getting a cue from the formatting that the spreadsheet was well formatted, turned off further judgmental processing.

Table 1: Error Rates in Spreadsheet Development

| Study | Spreadsheets | Percent of Spreadsheets Containing at Least One Error | Cell Error Rate (CER): Percent of Cells Containing Errors |
|---|---|---|---|
| *Field Audits* | | | |
| Davies and Ikin (1) | 19 | 21% | |
| Cragg & King | 20 | 25% | |
| Hicks (1995) | 1 | 100% | 1.2% |
| Butler (2) | 273 | 11% | |
| Coopers & Lybrand (3) | 23 | 91% | |
| KPMG (4) | 22 | 91% | |
| Lukasic | 2 | 100% | 2.2%, 2.5% |
| | | | |
| *Development Experiments* | | | |
| Brown & Gould | 27 | 63% | |
| Olson & Nilsen (5,6) | 14 | | 21% |
| Lerch (5,6) | 21 | | 9.3% |
| Hassinen (6) | 92 | 55% | 4.3% |
| Janvrin & Morrison (7) | 61 | | 6.6%-9.6% |
| Janvrin & Morrison (7) | | | 8.4%-16.8% |
| Kreie (posttest) | 73 | 42% | 2.5% |
| Panko & Halverson (1997) | 42 | 79% | 5.6% |
| Panko & Halverson (1998) | 35 | 86% | 4.6% |
| Panko & Sprague (8) | 26 | 35% | 2.1% |
| Panko & Sprague (9) | 17 | 24% | 1.1% |
| Teo & Tan | 168 | 42% | 2.1% |

(1) Only serious errors. (2) Only errors large enough to demand additional tax payments. (3) A value was off by at least 5%. (4) Only major errors. (5) Counted errors even if they were corrected. (6) CER is based only on formula cells. (7) CER is based only on high-risk formula cells. (8) MBA students with little or no development experience. (9) MBA students with at least 250 hours of spreadsheet development experience.

Source: Panko [2001b]. References to studies are given at the website.

Figure 1: Research Model

| Development Stage | Accuracy | Confidence | Interaction |
|---|---|---|---|
| Initial State | Not Applicable | Should depend on feedback from past testing or other sources of information. If no feedback or other information, likely to be higher than if feedback or other information (manipulation). (H1) | |
| During Development | Developer will make errors; some will be corrected | Will discover some errors during development and grow more confident. | May work more carefully if have better feedback on error rates. |
| After Development | Will contain errors not detected | Confidence should be higher than before development because of error discovery during development. (H2)<br><br>Confidence should depend on feedback from past testing or on other information (manipulation). (H3, H4) | May have made fewer errors with feedback or other information (the manipulation). (H5) |
| During Testing | Most errors will be discovered and fixed | The discovery of undetected errors during code inspection should reduce confidence. | Will be more likely to do after-development testing if had better feedback initially. |
| After Testing | Will contain fewer errors | Confidence should be lower because of feedback about the reality that people frequently do make errors. | May contain fewer errors. |

Table 2: Overconfidence in Monadic and Triadic Spreadsheet Development

| | Subjects working alone | | | Subjects working in triads | | |
|---|---|---|---|---|---|---|
| | Own alone | Other alone | All triads | Own triad | All alone | Other triads |
| N | 35 | 35 | 35 | 45 | 44 | 44 |
| Mean estimated probability of an error | 18% | 22% | 10% | 13% | 33% | 21% |
| Median estimated probability of an error | 10% | 10% | 5% | 5% | 25% | 10% |
| Actual percent of spreadsheets with errors | 86% | 86% | 27% | 27% | 86% | 27% |
| Percent of subjects who were overconfident | 100% | 100% | 97% | 86% | 100% | 80% |
| p | <0.0001 | <0.0001 | <0.0001 | <0.0001 | <0.0001 | 0.052 |

## Table 3: Results of the Warning Experiment

| Variable | Form A No Warning | Form B Warning (80% Kooker) (40% Wall) | P | Hypothesis |
|---|---|---|---|---|
| Number of Subjects | 27 | 28 | | |
| Both Tasks | | | | |
|   Both Incorrect | 93% | 75% | **0.040** | **H5** |
|   EEP Before: You (Averaged) | 37% | 56% | **0.008** | **H2** |
|   EEP After: You (Averaged) | 31% | 43% | 0.070 | H3 |
|   EEP After: Others (Averaged) | 23% | 40% | **0.001** | **H4** |
|   Decline in EEP Pre-Post | | | **0.0000** | **H2** |
| Kooker Task (more difficult) | | | | |
|   Actual | 85% | 68% | 0.068 | |
|   EEP Before: You | 40% | 60% | 0.011 | |
|   EEP After: You | 30% | 52% | 0.008 | |
|   EEP After: Others | 26% | 45% | 0.002 | |
|   Decline in EEP Pre-Post | | | 0.000 | |
| Wall Task (less difficult) | | | | |
|   Actual | 52% | 46% | 0.347 | |
|   EEP Before: You | 34% | 53% | 0.012 | |
|   EEP After: You | 31% | 34% | 0.344 | |
|   EEP After: Other | 20% | 35% | 0.007 | |
|   Decline in EEP Pre-Post | | | 0.000 | |

Note: EEP is the expected error probability: the subject's expressed likelihood that they or other developers would or did commit an error during development. The EEP for both tasks was the subject's average EEP for the two tasks.